\documentclass[twocolumn,preprintnumbers,amsmath,amssymb]{revtex4}
\usepackage{graphicx}
\usepackage{dcolumn}
\usepackage{bm}

\usepackage{amsthm}
\usepackage{epsfig}
\usepackage{subfigure}
\usepackage{amsmath}
\usepackage{indentfirst}
\usepackage{appendix}

\begin{document}

\preprint{}
\title{Rational solutions to multicomponent Yajima-Oikawa systems: from two dimension to one dimension  }
\author{Junchao Chen$^{1,2}$, Yong Chen$^{1}$\footnote{email: ychen@sei.ecnu.edu.cn}, Bao-Feng Feng$^{2}$\footnote{email: feng@utpa.edu}, Ken-ichi Maruno$^{3}$\footnote{email: kmaruno@waseda.jp}}
\affiliation{ $^{1}$Shanghai Key Laboratory of Trustworthy Computing, East China Normal University, Shanghai, 200062, People's Republic of China}
\affiliation{ $^{2}$Department of Mathematics, The University of Texas-Pan American, Edinburg, TX 78541, USA}
\affiliation{ $^{3}$Department of Applied Mathematics, School of Fundamental Science and Engineering,
Waseda University, 3-4-1 Okubo, Shinjuku-ku, Tokyo 169-8555, Japan}
\date{\today}

\begin{abstract}

Exact explicit rational solutions of two- and one- dimensional multicomponent Yajima-Oikawa (YO) systems, which contain multi-short-wave components and single long-wave one, are presented by using the bilinear method.
For two-dimensional system, the fundamental rational solution first describes the localized lumps, which have three different patterns: bright, intermediate and dark states.
Then, rogue waves can be obtained under certain parameter conditions and their behaviors are also classified to above three patterns with
different definition.
It is shown that the simplest (fundamental) rogue waves are line localized waves which arise from the constant background with
a line profile and then disappear into the constant background again.
In particular, two-dimensional intermediate and dark counterparts of rogue wave are found with the different parameter requirements.
We demonstrate that multirogue waves describe the interaction of several fundamental rogue waves, in which interesting curvy wave patterns appear in the intermediate times.
Different curvy wave patterns form in the interaction of different types fundamental rogue waves.
Higher-order rogue waves exhibit the dynamic behaviors that the wave structures start from lump and then retreats back to it, and this transient
wave possesses the patterns such as parabolas.
Furthermore, different states of higher-order rogue wave result in completely distinguishing lumps and parabolas.
Moreover, one-dimensional rogue wave solutions with three states are constructed through the further reduction.
Specifically, higher-order rogue wave in one dimensional case is derived under the parameter constraints.

PACS number(s): 05.45.Yv, 02.30.Ik, 47.35.Fg
\end{abstract}

\maketitle

\section{Introduction}

Rogue wave phenomena that ``appear from nowhere and
disappear without a trace \cite{akhmediev2009waves}", has recently become
one of the most active and important research areas
on both experimental observations and theoretical analysis,
since it exists in various different fields, including ocean \cite{kharif2009rogue},
optical systems \cite{solli2007optical,hohmann2010freak,montina2009non}, Bose--Einstein condensates \cite{bludov2009matter,bludov2010vector},
superfluids \cite{ganshin2008observation}, capillary waves \cite{shats2010capillary}, atmosphere \cite{stenflo2010rogue}, plasma \cite{moslem2011langmuir,bailung2011observation} and even in finance \cite{yan2011vector}.
From the mathematical description, rational solutions play a key role in the interpretation of the mechanisms underlying the formation and dynamics of rogue waves.
The first-order and most fundamental rational solution for nonlinear
Schr\"{o}dinger (NLS) equation was discovered by Peregrine \cite{peregrine1983water}.
Such a solution has the peculiarity of being localized in both space and time, and its maximum
amplitude reaches three times the constant background.
The hierarchy of higher-order rational solutions has been found \cite{akhmediev2009rogue,kedziora2012second,ankiewicz2011rogue,kedziora2011circular,dubard2010multi,dubard2011multi,gaillard2011families,guo2012nonlinear,ohta2012general}, in particular, in the framework of the integrable 1D NLS equation.
These higher-order waves were also localized in both coordinates, and could exhibit higher peak amplitudes or multiple intensity peaks.

Recently, apart from the NLS equation, exact rogue wave solutions have been
explored in a variety of nonlinear integrable systems such as the Hirota equation \cite{ankiewicz2010rogue,tao2012multisolitons}, the Sasa-Satsuma equation \cite{bandelow2012sasa,chen2013twisted} and the derivative NLS
equation \cite{xu2011darboux,guo2013high,chan2014rogue}.
More importantly, the relevant studies were also extended to coupled systems which usually involve
more than one component \cite{guo2011rogue,baronio2012solutions,bludov2010vector,zhao2013rogue,baronio2013rogue,chen2013rogue,chen2014dark,chen2014darboux,chen2014coexisting,
wang2014hihger,wang2014generalized,wang2015rogue}.
It was shown that compared with uncoupled systems, vector rogue wave solutions exhibit some novel structures
such as dark rogue wave.
In Ref. \cite{guo2011rogue,baronio2012solutions}, analytical rational solutions for the coupled NLS system allowed not only general vector
Peregrine soliton but also bright- and dark-rogue waves.

Moreover, the two-dimensional analogue of rogue wave, expressed by more complicated rational
form, have been recently reported in the Davey-Stewartson (DS)
equation \cite{ohta2012rogue,ohta2013dynamics} and Kadomtsev--Petviashvili-I equation \cite{dubard2011multi,dubard2013multi}.
In two kinds of DS systems \cite{ohta2012rogue,ohta2013dynamics},
the fundamental rogue waves are line rogue waves which arise from the constant background
and then retreat back to the constant background again.
More general rational solutions were divided into two categories: multi-rogue waves and higher order ones.
Multi-rogue waves describe the interaction between individual fundamental rogue waves,
whereas higher order rogue waves exhibit different dynamics such as the wavepacket rising from the constant
background but not decaying back to it.
Therefore, a natural
motivation is to investigate rational solutions in two-dimensional multicomponent system.
Specifically, it's worthy to expect appearance of a two-dimensional dark rogue wave counterpart, to our
best knowledge, which was never reported before.

Coming back to the one-dimensional case, rogue wave were usually obtained from homoclinic solutions by taking certain limits \cite{ankiewicz2010rogue,tao2012multisolitons,bandelow2012sasa,xu2011darboux,chan2014rogue,chen2013rogue}.
Indeed, most of literature devoted to the explicit expressions of rational solutions still resulted from the related homoclinic ones.
The construction of higher dimensional rational solutions may provide an alternative method for finding lower dimensional rogue wave through dimension reduction directly \cite{ohta2012rogue,ohta2013dynamics}.
In other words, one can generate the above rational solutions of
lower dimensional models from higher dimensional ones with the parameter constraints.
Application of reduction method to clarify the rational solution's relation between two different dimensions is also the aim of the present work.

In this paper, we focus on the two-dimensional multicomponent Yajima-Oikawa (YO) system, or the so-called 2D coupled
long-wave-short-wave resonance interaction system in which it comprises multi short-wave components and a single long-wave
component \cite{ohta2007two,kanna2009higher,kanna2013general}.
The long-wave--short-wave resonance interaction is a fascinating physical process in which a resonant interaction takes
place between a weakly dispersive long-wave and a short-wave
packet when the phase velocity of the former exactly
or almost matches the group velocity of the latter.
This phenomenon has been predicted
in plasma physics \cite{yajima1976formation,zakharov1972collapse}, nonlinear optics \cite{kivshar1992stable,chowdhury2008long} and hydrodynamics \cite{grimshaw1977modulation,djordjevic1977two,ma1979some}.
The rogue wave solutions to the 1D YO system had recently been derived by using Hirota bilinear method \cite{chow2013rogue} and Darboux transformation \cite{chen2014dark,chen2014darboux}. A special note of importance is that the coupled dark- and bright-field counterparts of the Peregrine soliton were demonstrated in \cite{chen2014dark,chen2014darboux,chen2014coexisting}.

The plan of the paper is as follows.
In Sec. II, we present exact and explicit rational solution for the two-dimensional multicomponent YO system by using the bilinear method.
In Sec. III,  dynamics of two-dimensional rational solution including fundamental lumps and general (multi- and higher-order) rogue waves are discussed in detail.
The one-dimensional rogue wave solution is derived through the further reduction and its dynamics are studied in Sec. IV.
The conclusion is given in the last section.

\section{ Explicit rational solution in the determinant form}

The two-dimensional multicomponent YO system:
\begin{subequations}\label{ryo-01}
\begin{eqnarray}
&&\textmd{i}(S^{(\ell)}_t + S^{(\ell)}_y) - S^{(\ell)}_{xx}+ L S^{(\ell)}=0,\ \ \ell =1,2,\cdots, M,\ \ \ \ \ \\
&&L_t=2\sum^M_{\ell=1}\sigma_\ell|S^{(\ell)}|^2_x,
\end{eqnarray}
\end{subequations}
where $\sigma_\ell=\pm1$, $S^{(\ell)}$ and $L$ indicate the $\ell$th short-wave and long-wave components, respectively.
When the wave propagation is independent of $y$ coordinate
Eq.(\ref{ryo-01}) is reduced to the one-dimensional multicomponent YO system

By the dependent variable transformation:
\begin{eqnarray}\label{ryo-02}
S^{(\ell)}= G^{(\ell)}_0 \frac{G^{(\ell)}}{F},\ \
 L= h-2\frac{\partial^2}{\partial x^2}\log F,
\end{eqnarray}
where $G^{(\ell)}_0=\rho_\ell\exp[{\rm i}(\alpha_\ell x+\beta_\ell y +\gamma_\ell t)]$, $ \gamma_\ell=h-\beta_\ell+\alpha^2_\ell$ and
$\alpha_\ell,\beta_\ell,\rho_\ell$ and $h$ are real parameters for $\ell =1,2,\cdots, M$,
the two-dimensional YO system can be cast into the bilinear form,
\begin{subequations}\label{ryo-03}
\begin{eqnarray}
&&\hspace{-1cm} [\textmd{i}(D_t+D_y-2\alpha_\ell D_x)-D^2_x]G^{(\ell)} \cdot F=0,\\
&&\hspace{-1cm} D_tD_xF \cdot F-2\sum^M_{\ell=1}\sigma_\ell F^2+2\sum^M_{\ell=1} \sigma_{\ell}\rho^2_{\ell}G^{(\ell)}G^{(\ell)*}=0,
\end{eqnarray}
\end{subequations}
where $F$ is a real variable, $G^{(\ell)}$ are complex variables, $*$ denotes the complex conjugation and $D$ is
Hirota's bilinear differential operator.

\emph{Theorem 1.} The two-dimensional multicomponent YO system has rational solution
(\ref{ryo-02}) with $F$ and $G^{(\ell)}$ given by $N \times N$ determinants
\begin{eqnarray}\label{ryo-04}
&& F=\tau'(n)\Big|_{n=0},\ \
 G^{(\ell)}=\tau'(n^{(\ell)}+1)\Big|_{n=0},
\end{eqnarray}
where $(n)\equiv (n^{(1)},n^{(2)},\cdots,n^{(M)})$,
$(n^{(\ell)}\pm1) \equiv (n^{(1)},n^{(2)},\cdots,n^{(\ell)}\pm1,\cdots,n^{(M)})$
and $n=0$ represents $n^{(1)}=n^{(2)}=\cdots n^{(\ell)} \cdots  =n^{(M)}=0$,
and the the matrix elements are defined by
\begin{eqnarray}\label{ryo-05}
&&  T'_{i,j}(n)=\prod^{M}_{\ell=1}(-\frac{p_i-{\rm i}\alpha_\ell}{p^*_j+{\rm i}\alpha_\ell})^{n^{(\ell)}}
\mathcal{A}_{i,j} \frac{1}{p_i+p^*_j}.
\end{eqnarray}
Here the operator $\mathcal{A}_{i,j}=\sum^{n_i}_{k=0}c_{ik}(\partial_{p_i}+\xi'_i+\sum^M_{\ell=1}\frac{n^{(\ell)}}{p_i-{\rm i}\alpha_\ell})^{n_i-k}  \sum^{n_j}_{l=0}c^*_{jl}(\partial_{p^*_j}+\xi'^*_j-\sum^M_{\ell=1}\frac{n^{(\ell)}}{p^*_j+{\rm i}\alpha_\ell})^{n_j-l}$ and
\begin{eqnarray}\label{ryo-06}
&& \xi'_i=-\sum^M_{\ell=1}\frac{\sigma_\ell{\rho^2_\ell}(t-y)}{(p_i-{\rm i}\alpha_\ell)^2}+x-2{\rm i}p_i y,
\end{eqnarray}
where $p_i$ and $c_{ik}$ are arbitrary complex constants, and $n_i$ is
an arbitrary positive integer.

It is emphasized that these rational solutions can also be expressed in term of Schur polynomials as shown in \cite{ohta2012rogue,ohta2013dynamics}.
Besides, from the appendix in \cite{ohta2012rogue,ohta2013dynamics}, one can know that the nonsingularity of rational solutions
exists if the real parts of wave numbers
$p_i$ ($1 \leqslant i \leqslant N$) are all positive or negative.

\section{Rational solutions for two-dimensional YO system}

In this section, we present the dynamics analysis of rational solutions to two-dimensional YO system in detail.

\subsection{Fundamental rational solutions}

As the simplest rational solution, one-rational solution
of first order is given by taking $N = 1$ and $n_1 = 1$,
\begin{widetext}
\begin{eqnarray}
\label{ryo-07}\nonumber   F &=& \sum^1_{k=0}c_{1k}(\partial_{p_1}+\xi'_1)^{1-k} \sum^1_{l=0}c^*_{1l}(\partial_{p^*_1}+\xi'^*_1)^{1-l} \frac{1}{p_1+p^*_1}\\
\nonumber &=& (\partial_{p_1} {+} \xi'_1+c_{11})(\partial_{p^*_1} {+} \xi'^{*}_1+c^*_{11}) \frac{1}{p_1+p^*_1}\\
 &=& \frac{1}{p_1+p^*_1} \Bigg[ \left(\xi'_1- \frac{1}{p_1+p^*_1}+c_{11}\right)
\left(\xi'^{*}_1 -\frac{1}{p_1+p^*_1}+c^*_{11} \right) + \frac{1}{(p_1+p^*_1)^2} \Bigg],\\
\label{ryo-08} \nonumber  G^{(\ell)} &=&  (-\frac{p_1-\textmd{i}\alpha_\ell}{p^*_1+\textmd{i}\alpha_\ell})  \sum^1_{k=0}c_{1k}(\partial_{p_1}+\xi'_1+\frac{1}{p_1-\textmd{i}\alpha_\ell})^{1-k}
\sum^1_{l=0}c^*_{1l}(\partial_{q_1}+\eta'_1-\frac{1}{p^*_1+\textmd{i}\alpha_\ell})^{1-l} \frac{1}{p_1+p^*_1} \\
\nonumber  &=& (-\frac{p_1-\textmd{i}\alpha_\ell}{p^*_1+\textmd{i}\alpha_\ell})
  (\partial_{p_1}+\xi'_1 +\frac{1}{p_1-\textmd{i}\alpha_\ell} +c_{11})
   (\partial_{q_1}+\xi'^{*}_1-\frac{1}{p^*_1+\textmd{i}\alpha_\ell} +c^*_{11}) \frac{1}{p_1+p^*_1}\\
  &=& (-\frac{p_1-\textmd{i}\alpha_\ell}{p^*_1+\textmd{i}\alpha_\ell})\frac{1}{p_1+p^*_1}
\Bigg[
\bigg(\xi'_1- \frac{1}{p_1+p^*_1}+c_{11}+\frac{1}{p_1-\textmd{i}\alpha_\ell} \bigg)
\bigg(\xi'^{*}_1  -\frac{1}{p_1+p^*_1}+ c^*_{11}-\frac{1}{p^*_1+\textmd{i}\alpha_\ell} \bigg)  +\frac{1}{(p_1+p^*_1)^2}
 \Bigg],\ \ \
\end{eqnarray}
\end{widetext}
with
\begin{eqnarray*}
&& \xi'_1=-\sum^M_{\ell=1}\frac{\sigma_\ell{\rho^2_\ell}(t-y)}{(p_1-{\rm i}\alpha_\ell)^2}+x-2{\rm i}p_1 y,\\
\end{eqnarray*}
where($c_{10}=1$) $p_1,c_{10}$ and $c_{11}$ are arbitrary complex constants.

Without loss of generality, we assume  $p_1=p_{1R}+{\rm i}p_{1I}$, $c_{11}=c_{11R}+{\rm i}c_{11I}$ and then rewrite above solution as
\begin{eqnarray*}
F&=&\frac{1}{p_1+p^*_1}(\theta_1\theta^*_1+\theta_0),\\
G^{(\ell)}&=&(-\frac{p_1-\textmd{i}\alpha_\ell}{p^*_1+\textmd{i}\alpha_\ell})\frac{1}{p_1+p^*_1}
[(\theta_1+\frac{1}{p_1-\textmd{i}\alpha_\ell})\\
&& \hspace{1cm}\times(\theta^*_1-\frac{1}{p^*_1+\textmd{i}\alpha_\ell})+\theta_0],\\
&=& (-\frac{p_1-\textmd{i}\alpha_\ell}{p^*_1+\textmd{i}\alpha_\ell})\frac{1}{p_1+p^*_1}
[(\theta_1+a_{1,\ell}+{\rm i}a_{2,\ell})\\
&&\hspace{1cm}\times(\theta^*_1-a_{1,\ell}+{\rm i}a_{2,\ell})+\theta_0],
\end{eqnarray*}
where
\begin{eqnarray*}
&&\hspace{-0.3cm} \theta_1=x + (b_1 + {\rm i}b_2)y + (c_1 + {\rm i}c_2)t  + d_1+{\rm i}d_2,\\
&&\hspace{-0.3cm} \theta_0=\frac{1}{(p_1+p^*_1)^2}=\frac{1}{4p^2_{1R}},\\
&&\hspace{-0.3cm} a_{1,\ell}=\frac{p_{1R}}{p^2_{1R}+(p_{1I}-\alpha_\ell)^2},\ \ a_{2,\ell}=\frac{\alpha_\ell-p_{1I}}{p^2_{1R}+(p_{1I}-\alpha_\ell)^2},\\
&&\hspace{-0.3cm} c_1=\sum^M_{\ell=1}\sigma_\ell\rho^2_\ell(a^2_{2,\ell}-a^2_{1,\ell}),\ \
 c_2=-2\sum^M_{\ell=1}\sigma_\ell\rho^2_\ell a_{1,\ell}a_{2,\ell}, \\
&&\hspace{-0.3cm} b_1=-c_1+2p_{1I},\ \ b_2=-c_2-2p_{1R},\\
&&\hspace{-0.3cm} d_1= -\frac{1}{2p_{1R}}+c_{11R},\ \ d_2=c_{11I}.
\end{eqnarray*}
Then, the final expression of the rational solutions reads
\begin{eqnarray}
\label{ryo-9} S^{(\ell)} &=& \bar{G}^{(\ell)}_0
\left[1-\frac{2{\rm i}(a_{1,\ell} l_2 -a_{2,\ell} l_1)+a^2_{1,\ell}+a^2_{2,\ell}}{l_1^2+l_2^2+\theta_0 } \right],\\
\label{ryo-10} L&=&h-4\frac{l^2_1-l^2_2-\theta_0}{(l_1^2+l_2^2+\theta_0)^2},
\end{eqnarray}
where $\bar{G}^{(\ell)}_0=-G^{(\ell)}_0 \frac{p_1-\textmd{i}\alpha_\ell}{p^*_1+\textmd{i}\alpha_\ell}$,
$l_1=x+b_1y+c_1t+d_1$ and $l_2=b_2y+c_2t+d_2$.

There are two different dynamical behaviors.

(i) Lump solution. When $b_2\neq 0$, one can see that the short-wave components $S^{(\ell)}$ and the long-wave component $L$ are constants along the $[x(t), y(t)]$ trajectory where
\begin{eqnarray*}
 x+b_1y+c_1t=0,\ \ b_2y+c_2t=0.
\end{eqnarray*}
Besides, at any fixed time, $(S^{(\ell)},L)\rightarrow (\bar{G}^{(\ell)}_0,h)$ when $(x,y)$ goes to infinity.
Hence these rational solutions are permanent lumps moving on the constant backgrounds.

Without loss of generality, the different patterns of lump solution can be discussed at time $t=0$ (after the shift of $t$).
In this situation, when $p_{1I}\neq\alpha_\ell$, the square of the short-wave amplitude $|S^{(\ell)}|^2$ possesses critical points
\begin{eqnarray}
\hspace{-0.8cm}&&(x_{1},y_{1})=\left(\frac{b_1}{b_2}d_2-d_1,-\frac{d_2}{b_2}\right),\\
\hspace{-0.8cm}&&(x^{(\ell)}_2,y^{(\ell)}_2)=\left({-}\frac{a_{2,\ell}}{a_{1,\ell}}(\mu_{\ell,1}{+}d_2)-\frac{b_1}{b_2}\mu_{\ell,1}-d_1,\frac{\mu_{\ell,1}}{b_2}\right),\ \ \\
\hspace{-0.8cm}&&(x^{(\ell)}_3,y^{(\ell)}_3)=\left(\frac{a_{1,\ell}}{a_{2,\ell}}(\mu_{\ell,2}{+}d_2)-\frac{b_1}{b_2}\mu_{\ell,2}-d_1,\frac{\mu_{\ell,2}}{b_2}\right),\ \
\end{eqnarray}
with
\begin{eqnarray*}
&&\mu_{\ell,1}=-c_{11I}\pm\frac{\sqrt{\Delta_{\ell,1}}}{2\Delta_{\ell,0}},\\ &&\mu_{\ell,2}=-c_{11I}\pm\frac{\sqrt{-(p_{1I}-\alpha_\ell)^2\Delta_{\ell,2}}}{2p_{1R}\Delta_{\ell,0}},\\
&& \Delta_{\ell,0}=(p_{1I}-\alpha_\ell)^2+p^2_{1R},\\
&& \Delta_{\ell,1}=3(p_{1I}-\alpha_\ell)^2-p^2_{1R},\\
&&\Delta_{\ell,2}=(p_{1I}-\alpha_\ell)^2-3p^2_{1R}.
\end{eqnarray*}
Note that $(x^{(\ell)}_3,y^{(\ell)}_3)$ are also two characteristic points. At these points, the second derivatives are
\begin{eqnarray}
\nonumber && H_1(\tilde{x},\tilde{y})=\frac{\partial^2|S^{(\ell)}|^2}{\partial x^2}\Bigg|_{(\tilde{x},\tilde{y})},\\
&& H_1(x_1,y_1)=\frac{192p^4_{1R}\Delta_{\ell,3}}{\Delta_{\ell,0}^2},\\
&&H_1(x^{(\ell)}_2,y^{(\ell)}_2)=-\frac{6p^4_{1R}\Delta_{\ell,0}}{(p_{1I}-\alpha_\ell)^4},\\
&&H_1(x^{(\ell)}_3,y^{(\ell)}_3)=6\Delta_{\ell,0},
\end{eqnarray}
with $\Delta_{\ell,3}=(p_{1I}-\alpha_\ell)^2-p^2_{1R}$ and the local quadratic forms are
\begin{eqnarray}
\hspace{-0.8cm}\nonumber  && H(\tilde{x},\tilde{y})=\left[\frac{\partial^2|S^{(\ell)}|^2}{\partial x^2}\frac{\partial^2|S^{(\ell)}|^2}{\partial y^2}-\left(\frac{\partial^2|S^{(\ell)}|^2}{\partial x \partial y}\right)^2\right]\Bigg|_{(\tilde{x},\tilde{y})},\\
\hspace{-0.8cm} && H(x_1,y_1)=\frac{4096b^2_2p^8_{1R}\Delta_{\ell,1}\Delta_{\ell,2}}{\Delta^4_{\ell,0}},\\
\hspace{-0.8cm} && H(x^{(\ell)}_2,y^{(\ell)}_2)=\frac{16b^2_2p^8_{1R}\Delta^2_{\ell,0}\Delta_{\ell,1}}{(p_{1I}-\alpha_\ell)^{10}},\\
\hspace{-0.8cm} && H(x^{(\ell)}_3,y^{(\ell)}_3)=\frac{16b^2_2\Delta^2_{\ell,0}\Delta_{\ell,1}}{p^2_{1R}}.
\end{eqnarray}
For one special case $p_{1I}=\alpha_\ell$, there are three critical points $(x_1,y_1)$ and $(x_4,y_4)=(\pm\frac{\sqrt{3}}{2p_{1R}}+\frac{b_1}{b_2}d_2-d_1,-\frac{d_2}{b_2})$.
Furthermore, $H_1(x_1,y_1)=-192p^2_{1R}, H(x_1,y_1)=12288p^4_{1R}b^2_2$, $H_1(x_4,y_4)=6p^2_{1R}, H(x_4,y_4)=48p^4_{1R}b^2_2$ and $|S^{(\ell)}|\Big|_{(x_4,y_4)}=0$.
Thus the lump solutions can be classified into three patterns:

(a) Bright state.  $0\leqslant(p_{1I}-\alpha_\ell)^2\leqslant\frac{1}{3}p^2_{1R}$: one local maximum, two characteristic points (when the sign take ``=", two local minimums located at two characteristic points);

(b) Intermediate state. $\frac{1}{3}p^2_{1R}<(p_{1I}-\alpha_\ell)^2<3p^2_{1R}$: two local maximums, two characteristic points;

(c) Dark state. $(p_{1I}-\alpha_\ell)^2\geqslant3p^2_{1R}$: two local maximums, one local minimum (when the sign take ``=", the local minimum located at the characteristic point).

\begin{figure}[!htbp]
{\includegraphics[height=1in,width=3.4in]{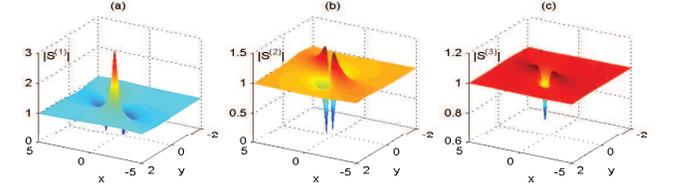}}
\caption{  One-lump for two-dimensional YO system with the parameters $(\sigma_1,\sigma_2,\sigma_3)=(1,1,1), (\rho_1,\rho_2,\rho_3)=(1,1,1)$, $(\alpha_1,\alpha_2,\alpha_3)=(\frac{3}{4},-\frac{1}{2},-3)$, $c_{11}=0$ and $p_1=1+\frac{1}{2}{\rm i}$ at the time $t=0$. \label{onelump2d}}
\end{figure}

The single lump profiles for the short-wave components are given in Fig.\ref{onelump2d} for $M=3$.
Three components represent three different patterns of lump solution in above classification respectively.

(ii) Rogue wave solution. When $b_2 = 0$, namely,
\begin{eqnarray}
\sum^M_{\ell=1} \frac{\sigma_\ell\rho^2_\ell (\alpha_\ell-p_{1I})}{[p^2_{1R}+(p_{1I}-\alpha_\ell)^2]^2}-1=0,
\end{eqnarray}
the rational solution are line waves, which are distinctly different from the moving line solitons. As $t\rightarrow\pm\infty$, these line waves go to uniform constant
backgrounds; in the intermediate times, they approach their bigger amplitudes.
More precisely, $(S^{(\ell)},L)\rightarrow (G^{(\ell)}_0,h)$ when $(x,y)$ goes to infinity.
Beside, when $p_{1I}\neq\alpha_\ell$, the square of the short-wave amplitude $|S^{(\ell)}|^2$ have critical lines:
\begin{eqnarray}
&& L_1:\ \ t=-\frac{d_2}{c_2},\ \ y=-\frac{1}{b_1}(x+d_1-\frac{c_1}{c_2}d_2),\\
\nonumber && L^{(\ell)}_2: t=\frac{\mu_{\ell,1}}{c_2},\ \ y=-\frac{1}{b_1}[x+d_1+\frac{c_1}{c_2}\mu_{\ell,1} \\
&&\hspace{4cm} +\frac{a_{2,\ell}}{a_{1,\ell}}(\mu_{\ell,1}+d_2)],\\
\nonumber && L^{(\ell)}_3: t=\frac{\mu_{\ell,2}}{c_2},\ \ y=-\frac{1}{b_1}[x+d_1+\frac{c_1}{c_2}\mu_{\ell,2}\\
&&\hspace{4cm}-\frac{a_{2,\ell}}{a_{1,\ell}}(\mu_{\ell,2}+d_2)].
\end{eqnarray}
Here $L^{(\ell)}_3$ are also two characteristic lines.
When $p_{1I}=\alpha_\ell$, there are three critical lines $L_1$ and
\begin{eqnarray}
L_4: t=-\frac{d_2}{c_2},\ \ y=-\frac{1}{b_1}(x+d_1-\frac{c_1}{c_2}d_2\pm\frac{\sqrt{3}}{2p_{1R}}),
\end{eqnarray}
which are also two characteristic lines.
Thus these line waves have the characteristics: appears from nowhere and disappears with no trace,
hence they are defined as line rogue waves.
Further analysis show that the feature of rogue wave for the short-wave component is classified into three patterns, which are summarized in Table I.

\begin{widetext}
\vglue10pt
\begin{center}
\tabcolsep=8pt
\small
\renewcommand\arraystretch{1}
\begin{minipage}{15.5cm}{
\small{\bf Table 1 } One-rogue wave for the short-wave component $S^{(\ell)}$.}
\end{minipage}
\vglue10pt
\begin{tabular}{ c  c  c  c  c  }
\hline
\hline
{State} & Condition $((p_{1I}-\alpha_\ell)^2=\Upsilon_\ell)$ & Local maximum & Local minimum & Zero-amplitude  \rule{0pt}{0.5cm}\\
\hline
& $\Upsilon_\ell=0$ & $L_1$ & $L_4$ & $L_4$ \rule{0pt}{0.4cm}\\
Bright    & $0<\Upsilon_\ell<\frac{1}{3}p^2_{1R}$ & $L_1$ & $L^{(\ell)}_3$ & $L^{(\ell)}_3$ \rule{0pt}{0.3cm}\\
   & $\Upsilon_\ell=\frac{1}{3}p^2_{1R}$ & $L_1=L^{(\ell)}_2$ & $L^{(\ell)}_3$ & $L^{(\ell)}_3$ \rule{0pt}{0.3cm}\\
   & $\frac{1}{3}p^2_{1R}<\Upsilon_\ell<p^2_{1R}$ & $L_1,L^{(\ell)}_2$,$|S^{(\ell)}(L_1)|< |S^{(\ell)}(L^{(\ell)}_2)|$ & $L^{(\ell)}_3$ & $L^{(\ell)}_3$ \rule{0pt}{0.6cm}\\
Intermediate & $\Upsilon_\ell=p^2_{1R}$ & $L^{(\ell)}_2$ & $L^{(\ell)}_3$ & $L^{(\ell)}_3$ \rule{0pt}{0.3cm}\\
   & $p^2_{1R}<\Upsilon_\ell<3p^2_{1R}$ & $L^{(\ell)}_2$ & $L_1,L^{(\ell)}_3$ & $L^{(\ell)}_3$ \rule{0pt}{0.3cm}\\
Dark & $\Upsilon_\ell=3p^2_{1R}$ & $L^{(\ell)}_2$ & $L_1=L^{(\ell)}_3$ & $L_1=L^{(\ell)}_3$ \rule{0pt}{0.6cm}\\
& $\Upsilon_\ell>3p^2_{1R}$ & $L^{(\ell)}_2 $ & $L_1$ & no \rule{0pt}{0.3cm}\\
 \hline
 \hline
\end{tabular}
\end{center}
\vglue10pt
\end{widetext}

\begin{figure}[!htbp]
{\includegraphics[height=2.8in,width=3.4in]{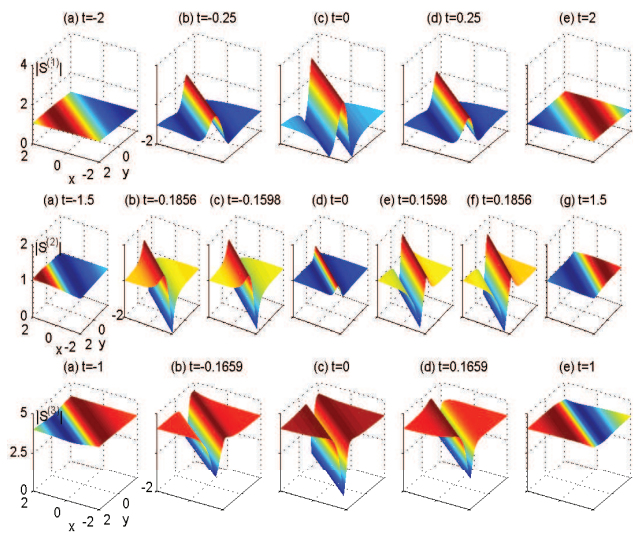}}
\caption{  One-rogue wave for two-dimensional YO system with the parameters $(\sigma_1,\sigma_2,\sigma_3)=(1,1,1)$, $(\rho_1,\rho_2,\rho_3)=(1,1,\frac{10}{7}\sqrt{6}+\frac{5}{14}\sqrt{2})$,
 $(\alpha_1,\alpha_2,\alpha_3)=(1,1-\frac{\sqrt{3}}{2},3)$, $c_{12}=0$ and $p_1=1+{\rm i}$. \label{onerogue2d}}
\end{figure}

Figure \ref{onerogue2d} displays one-rogue waves for the short-waves with $M=3$.
It can be clearly seen that three short-wave components describe different patterns of rogue wave as listed in Table I.
The amplitudes of $S^{(1)}$, $S^{(2)}$ and $S^{(3)}$ approach to the backgrounds $1$, $1$ and $4.0043$ respectively.
The component $S^{(1)}$ exhibits one bright rogue wave ($\Upsilon_1=0$), in which the amplitude attains its maximum $3$ at $L_1$ and minimum $0$ at $L_4$.
For the component $S^{(2)}$, as an example of intermediate state of rogue wave ($0<\Upsilon_2<\frac{1}{3}p^2_{1R}$),
its amplitude acquires the maximum $1.5275$ at $L^{(2)}_2$, the minimum $0$ at $L^{(2)}_3$,  and one local maximum $1.2857$ at $L_1$.
The amplitude of the component $S^{(3)}$ features a dark rogue wave ($\Upsilon_3>3p^2_{1R}$), which possesses the maximum $4.4770$ at $L^{(3)}_2$ and the minimum $0.8009$ at $L_1$.

From the above discussion on the one-lump and rogue wave for the short-wave components,
it is noted that the choice of the parameter $\alpha_\ell$ determines these local waves' patterns,
more specifically, the number, the position of extrema and zero point, and further the type of extrema of
the amplitude.
The same parameter's introduction is also carried out in the construction of dark--dark solitons for the coupled NLS system \cite{ohta2011general},
in which this treatment results in the generation of non-degenerate dark--dark solitons.

\subsection{Multi-rational solutions}

The multi-rational solutions can be obtained by taking $N>1$, $n_1=n_2
\cdots=n_N =1$ in the rational solution (\ref{ryo-05}).
These solutions describe the interaction
of $N$ individual fundamental rational solutions,
including lump and rogue wave, which depend on whether or not the parameters
meet the conditions
\begin{eqnarray}
\left\{ Im\left(f(p_i)\right)=0\Bigg| i=1,2,\cdots,N \right\},
\end{eqnarray}
where $Im$ represents the imaginary part of the function,
and $f(p_i)$ is defined by
\begin{eqnarray}
f(p_i)=\sum^M_{\ell=1}\frac{\sigma_\ell{\rho^2_\ell}}{(p_i-{\rm i}\alpha_\ell)^2}-2{\rm i}p_i.
\end{eqnarray}

\begin{figure}[!htbp]
{\includegraphics[height=2.8in,width=3.4in]{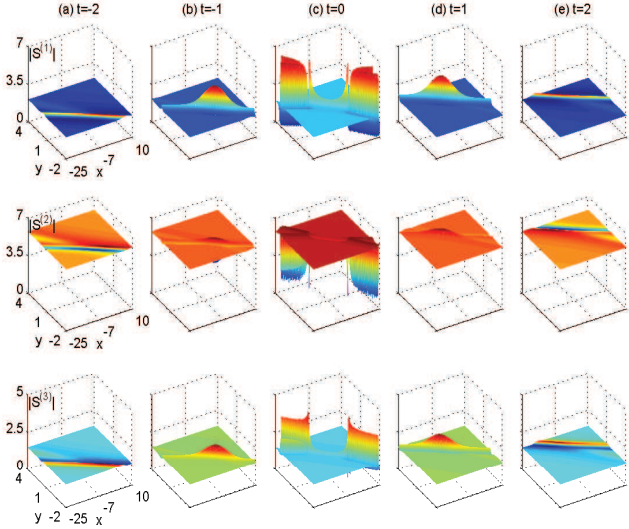}}
\caption{  Two-rogue wave for two-dimensional YO system with the parameters $(\sigma_1,\sigma_2,\sigma_3)=(1,-1,-1)$, $(\rho_1,\rho_2,\rho_3)=(2,6.0517,1.3600)$, $(\alpha_1,\alpha_2,\alpha_3)=(1,-1.6,1+\frac{\sqrt{2}}{2})$, $c_{11}=c_{21}=0$, $p_1=1+{\rm i}$ and $p_2=\frac{3}{2}+{\rm i}$. \label{tworogue2d}}
\end{figure}

For example, when $N=2$, one can write down $F$ and $G^{(\ell)}$ as
\begin{eqnarray}
F=\left|  \begin {array}{cc} T^0_{1,1} & T^0_{1,2}\\ T^0_{2,1} & T^0_{2,2} \end {array} \right|,\ \
G^{(\ell)}=\left|  \begin {array}{cc} T^1_{1,1} & T^1_{1,2}\\ T^1_{2,1} & T^1_{2,2} \end {array} \right|,
\end{eqnarray}
with
\begin{eqnarray*}
&& T^0_{i,j}=\frac{1}{p_i+p^*_j} \left[ \zeta_i \zeta'^*_j + \frac{1}{(p_i+p^*_j)^2} \right],\\
&& T^1_{i,j}=(-\frac{p_i-\textmd{i}\alpha_\ell}{p^*_j+\textmd{i}\alpha_\ell})\frac{1}{p_i+p^*_j}
\Bigg[
\left(\zeta_i+\frac{1}{p_i-\textmd{i}\alpha_\ell} \right) \\
&& \hspace{1.5cm} \times \left(\zeta'^{*}_j-\frac{1}{p^*_j+\textmd{i}\alpha_\ell} \right)  +\frac{1}{(p_i+p^*_j)^2}
 \Bigg],
\end{eqnarray*}
where $\zeta_i=\xi'_i- \frac{1}{p_i+p^*_j}+c_{i1}$, $\zeta'^{*}_j=\xi'^{*}_j -\frac{1}{p_i+p^*_j}+c^*_{j1}$,
$\xi'_i$ is given by (\ref{ryo-06}), $p_1$, $p_2$, $c_{11}$ and $c_{21}$ are arbitrary complex parameters.
In this case, the solution represent two-lump, two-rogue wave and the mixed solution consisting of one lump and one rogue wave by choosing different parameters. Here only two-rogue wave solutions are demonstrated in Fig. \ref{tworogue2d}.

\begin{figure}[!htbp]
{\includegraphics[height=2in,width=3.4in]{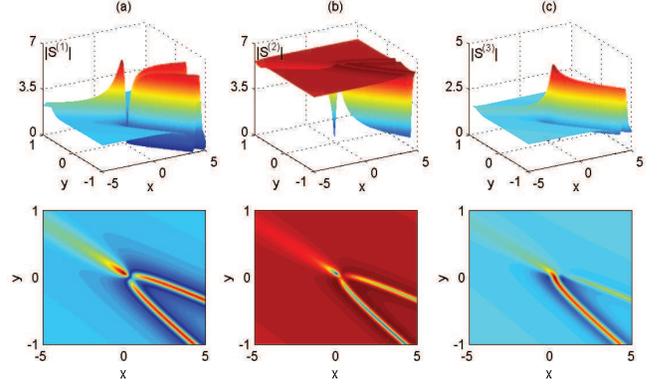}}
\caption{  Two-rogue wave for two-dimensional YO system with the same parameters as Fig. \ref{tworogue2d}. \label{tworogue2d-1}}
\end{figure}

As seen in Fig.\ref{tworogue2d}, the rogue wave of every short-wave component starts from the constant background in the entire $(x,y)$ plane (see the $t=-2$
panel).
In the intermediate times,
all three components undergo the nearly same process in which two line rogue waves interact with each other:
the regions of their intersection acquire higher/lower amplitudes first (see the $t=-1$
panel), the wave
patterns form into two curvy wave fronts which are completely separated (see the $t=0$
panel),
and then these waves possess higher/lower amplitudes again in the regions of their intersection (see the $t=1$
panel).
At large times, these solutions go back to the constant background (see the $t=2$
panel).

More interestingly, three short-wave components display the rogue wave with
bright-bright state $(p_{1I}-\alpha_1=0$ and $p_{2I}-\alpha_1=0$ for $S^{(1)})$, dark-dark state $((p_{1I}-\alpha_2)^2>3p^2_{1R}$ and $(p_{2I}-\alpha_2)^2>3p^2_{1R}$ for $S^{(2)})$ and intermediate-bright state $(\frac{1}{3}p^2_{1R}<(p_{1I}-\alpha_3)^2<p^2_{1R}$ and $0<(p_{2I}-\alpha_3)^2<\frac{1}{3}p^2_{1R}$ for $S^{(3)})$.
An inspection of the right branch of two curvy wave fronts ($t=0$) in the small region (see Fig.\ref{tworogue2d-1})
indicates that three are different features for every components.
At the vertices of three curvy wave fronts, $S^{(1)}$ with
bright-bright state has one humped hole, $S^{(2)}$ with
dark-dark state has one sunken hole and $S^{(3)}$ with intermediate-bright state has one humped hole and two sunken holes.
As the detailed analysis for the pattern of single rogue wave,  such two-rogue wave structures are controlled collectively by the complex parameter $p_i$ and the real parameter $\alpha_\ell$.

\subsection{Higher-order rational solutions}

The higher order rational solutions can be obtained by taking $N=1$ and $n_1>1$ in the rational solution (\ref{ryo-05}).
In this situation, these solutions are viewed as higher-order lumps and rogue waves.
Notice that if the parameters satisfy the following relations:
\begin{eqnarray}\label{highcs}
\left\{ Im\left(\frac{d^kf(p_1)}{dp^k_1}\right)=0\Bigg| k=0,1,2,\cdots,\tilde{N} \right\},
\end{eqnarray}
the imaginary part of the coefficient of $y$ will be zero.
In such a special case, one can get the $\tilde{N}$-order rogue waves solutions.

For instance, if $n_1 = 2$, the functions $F$ and $G^{(\ell)}$ take the form $(c_{11}=0)$
\begin{eqnarray}
 && \hspace{-0.5cm} F =  [(\partial_{p_1} {+} \xi'_1)^2+c_{12}][(\partial_{p^*_1} {+} \xi'^{*}_1)^2+c^*_{12}] \frac{1}{p_1+p^*_1},\\
\nonumber &&\hspace{-0.85cm} G^{(\ell)}  = (-\frac{p_1-\textmd{i}\alpha_\ell}{p^*_1+\textmd{i}\alpha_\ell})
  [(\partial_{p_1}+\xi'_1 +\frac{1}{p_1-\textmd{i}\alpha_\ell})^2 +c_{12}]\\
  &&\hspace{0.6cm}\times [(\partial_{q_1}+\xi'^{*}_1-\frac{1}{p^*_1+\textmd{i}\alpha_\ell})^2 +c^*_{12}] \frac{1}{p_1+p^*_1},
\end{eqnarray}
where $\xi_1$ is defined by (\ref{ryo-06}), $p_1$ and $c_{12}$ are arbitrary complex parameters.
For the general choice of the parameters, this solution represent second-order lump.
When these parameters meet the constraint conditions (\ref{highcs}) with $\tilde{N}=2$,
this rational solution reduces to second-order rogue wave solutions.

\begin{figure}[!htbp]
{\includegraphics[height=2.8in,width=3.4in]{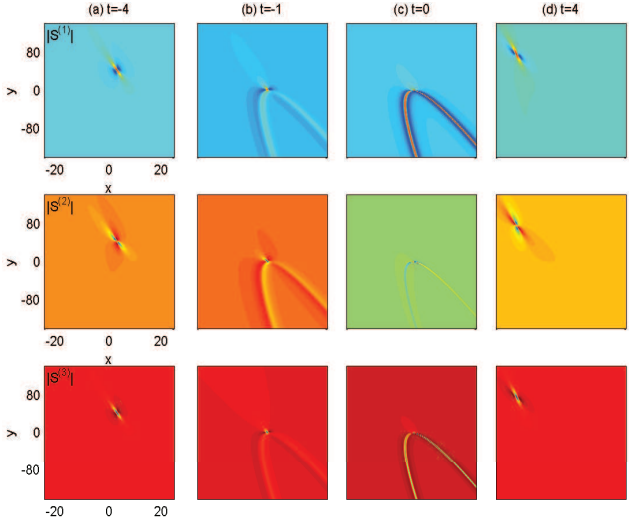}}
\caption{  Second-order rogue wave for two-dimensional YO system with the parameters $(\sigma_1,\sigma_2,\sigma_2)=(1,1,1)$, $(\rho_1,\rho_2,\rho_3)=(1,\frac{\sqrt{29}}{5},\frac{7\sqrt{6}}{3})$, $(\alpha_1,\alpha_2,\alpha_3)=(\frac{2}{3},2,4)$, $c_{12}=0$ and $p_1=1+{\rm i}$. \label{secondrogue2d}}
\end{figure}

In Fig.\ref{secondrogue2d}, we illustrate the second-order rogue wave for two dimensional YO system which still contains three short-wave components.
This kind of construction for higher-order rogue wave leads to a new phenomenon: these higher-order rogue waves do
not uniformly approach the constant background but feature localized lumps as $t\rightarrow\pm\infty$, which is different from the multirogue
waves discussed above.
As seen from Fig.\ref{secondrogue2d}, the solutions for three short-waves are localized lumps sitting on the
constant backgrounds when $|t|\geqslant1$ (see the $t=\pm4$ panels).
When $t\rightarrow 0$, these lumps disappear gradually and three parabola-shaped rogue waves rise from their backgrounds (see the $t=-1,0$ panels).
In addition, three short-waves exhibit three different patterns of rogue waves throughout the process of their shape change.
The solutions are second-order rogue waves with bright state for $S^{(1)}$ ($(p_{1I}-\alpha_2)^2<\frac{1}{3}p^2_{1R}$), intermediate state for $S^{(2)}$ ($(p_{1I}-\alpha_2)^2=p^2_{1R}$) and dark state for $S^{(3)}$ ($(p_{1I}-\alpha_2)^2>3p^2_{1R}$).
Visually, the components $S^{(1)}$, $S^{(2)}$ and $S^{(3)}$ undergo bright, intermediate and dark lumps at $t=\pm4$, and especially humped, sunken-humped and humped parabola fronts at $t=0$ respectively.

\section{Rational solutions for one-dimensional YO system}

Consider the further reduction, two-dimensional multi-component YO system becomes one-dimensional one.
Therefore the rational solutions for one-dimensional multi-component YO system can be derived
from ones of two-dimensional case. More specifically, the following theorem is summarized:

\emph{Theorem 2.} The one-dimensional multicomponent YO system:
\begin{subequations}
\begin{eqnarray}
&&\textmd{i}S^{(\ell)}_t  - S^{(\ell)}_{xx}+ L S^{(\ell)}=0,\ \ \ell =1,2,\cdots, M,\ \ \ \ \ \\
&&L_t=2\sum^M_{\ell=1}\sigma_\ell|S^{(\ell)}|^2_x,
\end{eqnarray}
\end{subequations}
where $\sigma_\ell=\pm1$, has rational solution
\begin{eqnarray}\label{one-sol}
S^{(\ell)}= G^{(\ell)}_0 \frac{G^{(\ell)}}{F},\ \
 L= h-2\frac{\partial^2}{\partial x^2}\log F,
\end{eqnarray}
where $G^{(\ell)}_0=\rho_\ell\exp[{\rm i}(\alpha_\ell x +\gamma_\ell t)]$, $ \gamma_\ell=h+\alpha^2_\ell$ and
$\alpha_\ell,\rho_\ell$ and $h$ are real parameters for $\ell =1,2,\cdots, M$.
Here $F$ and $G^{(\ell)}$ are defined by (\ref{ryo-04})-(\ref{ryo-06}) and the parameters
satisfy the constraints
\begin{eqnarray}\label{one-cs}
\left\{ \frac{d^kf(p_i)}{dp^k_1}=0\Bigg|i=1,2,\cdots,N; k=0,1,2,\cdots,\tilde{N} \right\}.
\end{eqnarray}

\subsection{Fundamental rational solution}
\begin{figure}[!htbp]
{\includegraphics[height=2in,width=3.4in]{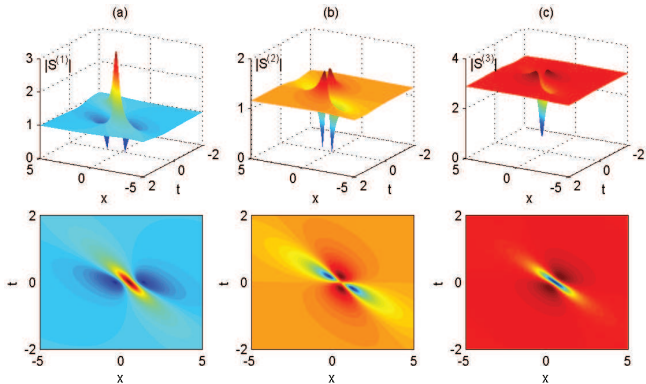}}
\caption{  One rogue wave for one-dimensional YO system with the parameters $c_{11}=0$, $(\sigma_1,\sigma_2,\sigma_3)=(-1,1,1)$, $(\alpha_1,\alpha_2,\alpha_3)=(1,2,3)$, $(\rho_1,\rho_2,\rho_3)=(1,\frac{2\sqrt{3}}{3},\frac{5\sqrt{3}}{3})$ and $p_1=1+{\rm i}$.\label{onerogue1d}}
\end{figure}

According to Theorem 2, the fundamental rational solution for one-dimensional multi-component YO system
has same form as Eqs.(\ref{ryo-9})-(\ref{ryo-10}) but the parameters need to meets the requirement (\ref{one-cs}) for $N=1$ and $k=0$, namely,
\begin{eqnarray}
f(p_1)=0.
\end{eqnarray}
As reported in \cite{chen2014dark,chen2014darboux}, the rogue wave of the short-wave component can be classified into bright, intermediate, and dark
states.
Here, one can find that $(S^{(\ell)},L)$ still approaches  $(\bar{G}^{(\ell)}_0,h)$ as $(x,t)\rightarrow \pm\infty$.
Meanwhile, when $p_{1I}\neq\alpha_\ell$, the square of the short-wave amplitude $|S^{(\ell)}|^2$ possesses critical points
\begin{eqnarray}
\hspace{-0.8cm}&&(x_{1},t_{1})=\left(\frac{c_1}{c_2}d_2-d_1,-\frac{d_2}{b_2}\right),\\
\hspace{-0.8cm}&&(x^{(\ell)}_2,t^{(\ell)}_2)=\left({-}\frac{a_{2,\ell}}{a_{1,\ell}}(\mu_{\ell,1}{+}d_2)-\frac{c_1}{c_2}\mu_{\ell,1}-d_1,\frac{\mu_{\ell,1}}{c_2}\right),\ \ \\
\hspace{-0.8cm}&&(x^{(\ell)}_3,t^{(\ell)}_3)=\left(\frac{a_{1,\ell}}{a_{2,\ell}}(\mu_{\ell,2}{+}d_2)-\frac{c_1}{c_2}\mu_{\ell,2}-d_1,\frac{\mu_{\ell,2}}{c_2}\right),\ \
\end{eqnarray}
and characteristic points $(x^{(\ell)}_3,t^{(\ell)}_3)$.
When $p_{1I}=\alpha_\ell$, there are three critical points $(x_1,t_1)$ and $(x_4,t_4)=(\pm\frac{\sqrt{3}}{2p_{1R}}+\frac{c_1}{c_2}d_2-d_1,-\frac{d_2}{c_2})$ and characteristic points $(x_4,t_4)$.
Then the detailed calculation show that the domains for three states are $0\leqslant(p_{1I}-\alpha_\ell)^2\leqslant\frac{1}{3}p^2_{1R}$ (when the sign take ``=", two local minimums located at two characteristic points),
$\frac{1}{3}p^2_{1R}<(p_{1I}-\alpha_\ell)^2<3p^2_{1R}$ and $(p_{1I}-\alpha_\ell)^2\geqslant3p^2_{1R}$ (when the sign take ``=", the local minimum located at the characteristic point),  respectively.
Three different kind of rogue wave structures are demonstrated for every short-wave components in Fig. \ref{onerogue1d}.

In Ref.\cite{ohta2012rogue,ohta2013dynamics}, Ohta \emph{et al.} have shown that the fundamental rogue waves in the DS equations are
two-dimensional counterparts of the fundamental (Peregrine) rogue waves in the NLS equation.
Meanwhile, Dubard \emph{et al.} have created two-dimensional rogue waves via the NLS-KP correspondence \cite{dubard2011multi,dubard2013multi}.
Very similarly, for the YO system, the two-dimensional rogue waves are viewed as the counterparts of one-dimensional ones.
From above discussion, one can find such a fact that by further taking the real parts of functions $f(p_i)$ as zero, two-dimensional rogue waves is reduced to one-dimensional ones,
or by restricting $f(p_i)=0$, one-dimensional rogue waves can be acquired from two-dimensional lump solutions.

\subsection{Nonfundamental rogue wave}

\begin{figure}[!htbp]
{\includegraphics[height=2in,width=3.4in]{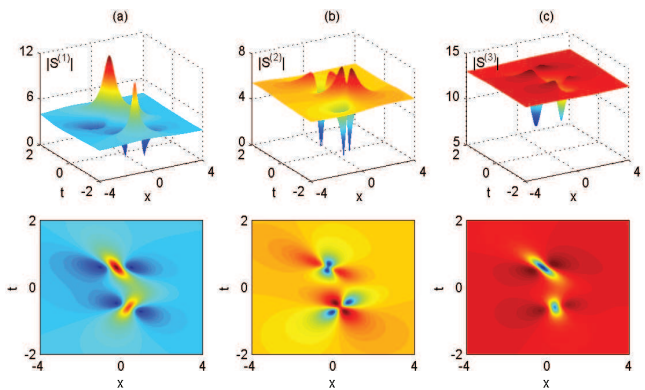}}
\caption{  Second-order rogue wave for one-dimensional YO system with the parameters
$c_{12}=1$, $(\sigma_1,\sigma_2,\sigma_3)=(1,1,1)$, $(\alpha_1,\alpha_2,\alpha_3)=(0.5,-1,3.7839)$, $(\rho_1,\rho_2,\rho_3)=(3.9698,5.2657,13.0687)$ and $p_1=1.25+0.25{\rm i}$.\label{secondrogue1d}}
\end{figure}

\begin{figure}[!htbp]
{\includegraphics[height=1in,width=3.4in]{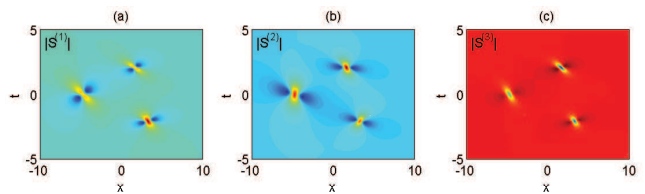}}
\caption{  Third-order rogue wave for one-dimensional YO system with the parameters
$(\sigma_1,\sigma_2,\sigma_3)=(1,1,1)$, $(\alpha_1,\alpha_2,\alpha_3)=(-0.4109,0.5044,3)$, $(\rho_1,\rho_2,\rho_3)=(1.7757,1.5887,6.2015)$, $c_{13}=150,c_{11}=c_{12}=0$ and $p_1=1.0218+0.2{\rm i}$.\label{thirdrogue1d}}
\end{figure}

As in the two-dimensional case, here one can consider two subclasses of these nonfundamental
rogue waves: multi- and higher order rogue waves. Specifically, by restricting the parameters
\begin{eqnarray}
\left\{ f(p_i)=0\Bigg| i=1,2,\cdots,N \right\},
\end{eqnarray}
in Theorem 2, (\ref{one-sol}) is identified as multi-rational solution
and by imposing the constraint conditions
\begin{eqnarray}
\left\{ \frac{d^kf(p_1)}{dp^k_1}=0\Bigg| k=0,1,2,\cdots,\tilde{N} \right\},
\end{eqnarray}
in Theorem 2, the higher-order rational solution can be written.
For illustrative purpose, we only present plots of second- and third-order rogue waves for the short-wave components in Figs.\ref{secondrogue1d} and \ref{thirdrogue1d}.
It can be observed from Fig.\ref{secondrogue1d} that second-order short-wave solutions contain bright-bright, intermediate-intermediate and dark-dark rogue waves.
Fig.\ref{thirdrogue1d} depicts third-order rogue waves for the short-wave components, in which three components describe the nonfundamental rogue wave with different mixed states respectively.

\section{Conclusion}

In conclusion, we derive exact explicit rational solutions of two- and one- dimensional multicomponent Yajima-Oikawa systems consisting of multi-short-wave components and single long-wave one.
These solutions in terms of determinants are obtained by using the bilinear method.
In two-dimensional case, the fundamental rational solution first describes the localized lumps,
which have three different patterns: bright, intermediate and dark states.
Further, by inserting certain parameter constraint conditions,
rogue waves can be reduced from the general rational solutions.
Rogue waves behaviors were also classified to above three patterns but with different description.
We show that the simplest (fundamental) rogue waves are line localized waves
which arise from the constant background with a line profile and then disappear into the constant background again.
In particular, we report two-dimensional intermediate and dark counterparts of rogue wave by considering the different parameter requirements.
Two subclasses of nonfundamental rogue waves, multi- and higher-order ones are discussed in detail.
Multirogue waves describe the interaction of several fundamental rogue waves,
in which interesting curvy wave patterns appear in the intermediate times.
Meanwhile, in the interaction of different types fundamental rogue waves, the corresponding different curvy wave patterns occur.
Higher-order rogue waves exhibit the dynamic behaviors that the wave structure start from lump and then retreats back to it, and this transient wave possesses patterns such as parabolas.
Furthermore, different states of higher-order rogue wave result in completely distinguishing lumps and parabolas.
In addition, by considering the further reduction, one-dimensional rogue wave solutions with three states are constructed.
Specifically, higher-order rogue wave in one dimensional case is derived under the parameter constraints.
Ours results, especially two-dimensional intermediate and dark counterparts of rogue wave,
are expected to be a crucial progress in the physical understanding of higher-dimensional rogue waves in the fields such as oceanography and nonlinear optics.

Finally, apart from the existence of the vector rogue waves,
the family of semirational vector solution for coupled NLS equations reported in Ref.\cite{guo2011rogue,baronio2012solutions}
also described a kind of interaction wave among rogue wave and other localized waves including periodic breather and soliton.
Therefore, one can investigate the similar dark-bright boomeronic solitons in the multicomponent YO system.
More importantly,
the study of such solutions can be extend to the higher-dimensional multicomponent coupled systems.
The corresponding semirational solution may provide evidence of an interesting interaction
between the dark-bright boomeronic solitons and the rogue
wave in higher-dimensional situation.
The KP hierarchy reduction method, which has used to derive the two-dimensional counterparts of rogue wave by Ohta and Yang \cite{ohta2012rogue,ohta2013dynamics}, can be applied to attain
the general semirational solution for the two-dimensional YO equation and other higher-dimensional coupled integrable systems.
We will report the relevant results elsewhere.

\section*{Acknowledgments}
The project is supported by the Global Change Research Program of China
(No.2015CB953904), National Natural Science Foundation of China (Grant No.
11275072 and 11435005), Research Fund for the Doctoral Program of Higher
Education of China (No. 20120076110024), Innovative Research Team Program of
the National Natural Science Foundation of China (Grant No. 61321064),
Shanghai Knowledge Service Platform for Trustworthy Internet of Things under
Grant No. ZF1213, Shanghai Minhang District talents of high level scientific
research project, Talent Fund and K.C. Wong Magna Fund in Ningbo
University.

\section*{Appendix}
\setcounter{equation}{0}
 \renewcommand\theequation{A\arabic{equation}}

In this appendix, we will prove Theorem 1 in Sec. II by
using the bilinear method. First we present the following
lemma:

\emph{Lemma 1}
The bilinear equations in the KP hierarchy:
\begin{subequations}\label{ryo-37}
\begin{eqnarray}
&&\hspace{-1cm} (D_{x_2}-D^2_{x_1}-2a_\ell D_{x_1})\tau(n^{(\ell)}{+}1) \cdot \tau(n)=0,\\
&&\hspace{-1cm} (\frac{1}{2}D_{x_1}D_{r_\ell}-1)\tau(n)\cdot \tau(n) =-\tau(n^{(\ell)}{+}1)\tau(n^{(\ell)}{-}1),\ \ \
\end{eqnarray}
\end{subequations}
for $ \ell=1,2,\cdots,M$, where $(n)\equiv (n^{(1)},n^{(2)},\cdots,n^{(M)})$,
$(n^{(\ell)}\pm1) \equiv (n^{(1)},n^{(2)},\cdots,n^{(\ell)}\pm1,\cdots,n^{(M)})$,
$a_l$ are complex constants, and $n^{(\ell)}$ are integers,
have the Gram determinant solutions
\begin{subequations}\label{ryo-38}
\begin{eqnarray}
\tau(n)=\det_{1\leq i,j\leq N}\Big(T_{ij}(n)\Big)=\Big|T_{ij}(n)\Big|_{1\leq i,j\leq N},
\end{eqnarray}
with the matrix element
\begin{eqnarray}
&& T_{ij}(n)= \prod^{M}_{\ell=1}(-\frac{p_i-a_\ell}{q_j+a_\ell})^{n^{(\ell)}} \cdot \frac{1}{p_i+q_j} \textmd{e}^{\xi_i+\eta_j},\\
&& \xi_i=\sum^M_{\ell=1}\frac{1}{p_i-a_\ell}r_{\ell}  + p_ix_1 + p^2_ix_2 + \xi_{i0},\\
&& \eta_j=\sum^M_{\ell=1}\frac{1}{q_i+a_\ell}r_{\ell}  + q_jx_1 - q_j^2 x_2 + \eta_{j0}.
\end{eqnarray}
where $p_i,q_j,\xi_{i0}$ and $\eta_{j0}$ are complex constants.
\end{subequations}

In order to get rational solutions, by introducing the differential operators:
\begin{eqnarray}\label{ryo-39}
A_i=\sum^{n_i}_{k=0}c_{ik} \partial_{p_i}^{n_i-k},\ \ B_j=\sum^{n_j}_{l=0}d_{jl} \partial_{q_i} ^{n_j-l},
\end{eqnarray}
where $c_{ik},d_{jl}$ are arbitrary complex constants, and acting the matrix element $T_{ij}(n)$ in (\ref{ryo-38}),
the solutions
\begin{eqnarray}
\tau'(n)=\det_{1\leq i,j\leq N}\Big(T'_{ij}(n)\Big),\ \ T'_{ij}(n)=A_iB_jT_{ij}(n),
\end{eqnarray}
still satisfy the bilinear equations (\ref{ryo-37}).

By using the operator relations
\begin{subequations}
\begin{eqnarray}
&&\hspace{-0.5cm} \nonumber (\partial_{p_i}) \prod^M_{\ell=1} (p_i-a_\ell)^{n^{(\ell)}} \textmd{e}^{\xi_i} \\
&& =\prod^M_{\ell=1} (p_i-a_\ell)^{n^{(\ell)}} \textmd{e}^{\xi_i}[\partial_{p_i}+\xi'_i+ \sum^M_{\ell=1}\frac{n^{(\ell)}}{p_i-a_\ell}],\ \ \ \\
&&\hspace{-0.7cm} \nonumber (\partial_{q_j}) \prod^M_{\ell=1} (-q_j-a_\ell)^{-n^{(\ell)}}\textmd{e}^{\eta_j}\\
&&\hspace{-0.5cm} = \prod^M_{\ell=1} (-q_j-a_\ell)^{-n^{(\ell)}}   \textmd{e}^{\eta_j}
[\partial_{q_j}+\eta'_j-\sum^M_{\ell=1}\frac{n^{(\ell)}}{q_j+a_\ell}],
\end{eqnarray}
where
\begin{eqnarray}
&& \xi'_i=-\sum^M_{\ell=1}\frac{r_\ell}{(p_i-a_\ell)^2} +x_1+2p_ix_2,\\
&& \eta'_j=-\sum^M_{\ell=1}\frac{r_\ell}{(q_i+a_\ell)^2} +x_1-2q_jx_2,
\end{eqnarray}
\end{subequations}
the matrix element $T'_{i,j}(n)$ in (\ref{ryo-38}) becomes the following form
\begin{eqnarray}
 T'_{i,j}(n)&=&\prod^{M}_{\ell=1}(-\frac{p_i-a_\ell}{q_j+a_\ell})^{n^{(\ell)}} \textmd{e}^{\xi_i+\eta_j}
\bar{\mathcal{A}}_{i,j} \frac{1}{p_i+q_j},
\end{eqnarray}
where the operator $\bar{\mathcal{A}}_{i,j}=\sum^{n_i}_{k=0}c_{ik}(\partial_{p_i}+\xi'_i+\sum^M_{\ell=1}\frac{n^{(\ell)}}{p_i-a_\ell})^{n_i-k}  \sum^{n_j}_{l=0}d_{jl}(\partial_{q_j}+\eta'_j-\sum^M_{\ell=1}\frac{n^{(\ell)}}{q_j+a_\ell})^{n_j-l}$.

Further, taking parameter constraints
\begin{eqnarray}
&& a_\ell=\textmd{i}\alpha_\ell,  \ \ p_i =q^*_i, \ \ d_{jl}=c^*_{jl},\ \ \xi_{i0}=\eta^*_{i0}
\end{eqnarray}
and assuming $r_\ell,x_1$ are real, $x_2$ are pure imaginary,  we have
\begin{eqnarray}
\eta_j=\xi^*_j, \ \  T'^{*}_{j,i}(n)=T'_{i,j}(-n),\ \ \tau'^{*}(n)=\tau'(-n).
\end{eqnarray}

Let
\begin{eqnarray*}
&& F=\tau'(n)\Big|_{n^{(1)}=n^{(2)}=\cdots n^{(\ell)} \cdots  =n^{(M)}=0},\\
&& G^{(\ell)}=\tau'(n^{(\ell)}+1)\Big|_{n^{(1)}=n^{(2)}=\cdots n^{(\ell)} \cdots  =n^{(M)}=0},\\
&& G^{(\ell)*}=\tau'(n^{(\ell)}-1)\Big|_{n^{(1)}=n^{(2)}=\cdots n^{(\ell)} \cdots  =n^{(M)}=0},
\end{eqnarray*}
Eqs.(\ref{ryo-37}) become
\begin{subequations}\label{ryo-a9}
\begin{eqnarray}
&& (D_{x_2}-2\textmd{i}\alpha_\ell D_{x_1}-D^2_{x_1})G^{(\ell)}\cdot F=0,\\
&& (\frac{1}{2}D_{x_1}D_{r_\ell}-1)F\cdot F=-G^{(\ell)}G^{(\ell)*},
\end{eqnarray}
\end{subequations}
for $\ell=1,2,\cdots,M$, and meanwhile the element of $\tau$ function is expressed by
\begin{eqnarray}
&&  T'_{i,j}(n)=\prod^{M}_{\ell=1}(-\frac{p_i-{\rm i}\alpha_\ell}{p^*_j+{\rm i}\alpha_\ell})^{n^{(\ell)}} \textmd{e}^{\xi_i+\xi^*_j}
\bar{\bar{\mathcal{A}}}_{i,j} \frac{1}{p_i+p^*_j},\ \ \ \ \ \
\end{eqnarray}
where the operator $\bar{\bar{\mathcal{A}}}_{i,j}=\sum^{n_i}_{k=0}c_{ik}(\partial_{p_i}+\xi'_i+\sum^M_{\ell=1}\frac{n^{(\ell)}}{p_i-{\rm i}\alpha_\ell})^{n_i-k}  \sum^{n_j}_{l=0}c^*_{jl}(\partial_{p^*_j}+\xi'^*_j-\sum^M_{\ell=1}\frac{n^{(\ell)}}{p^*_j+{\rm i}\alpha_\ell})^{n_j-l}$ and
\begin{eqnarray*}
&& \xi_i=\sum^M_{\ell=1}\frac{r_\ell}{p_i-\textmd{i}\alpha_\ell}  + p_ix_1 + p^2_ix_2 + \xi_{i0},\\
&& \xi^*_j=\sum^M_{\ell=1}\frac{r_\ell}{p^*_j+\textmd{i}\alpha_\ell} + p^*_j x_1 - p^{*2}_j x_2 + \xi^*_{j0},\\
&& \xi'_i=-\sum^M_{\ell=1}\frac{r_{\ell}}{(p_i-{\rm i}\alpha_\ell)^2}+x_1+2p_ix_2,\\
&& \xi'^*_j=-\sum^M_{\ell=1}\frac{r_{\ell}}{(p^*_j+{\rm i}\alpha_\ell)^2}+x_1-2p^*_jx_2.
\end{eqnarray*}

Applying the change of independent variables
\begin{subequations}
\begin{eqnarray}
  &&  x_1=x,\ \ x_2=-\textmd{i}y,\ \ r_\ell=\sigma_\ell{\rho^2_\ell}(t-y),
\end{eqnarray}
i.e.,
\begin{equation}
    \partial_x=\partial_{x_1},\ \ \partial_y= -\textmd{i}\partial_{x_2} -\sum^M_{\ell=1} \sigma_\ell{\rho^2_\ell}\partial_{r_\ell},\ \ \partial_t=\sum^M_{\ell=1} \sigma_\ell{\rho^2_\ell}\partial_{r_\ell},
\end{equation}
\end{subequations}
Eqs.(\ref{ryo-a9}) are reduced to the bilinear
equation (\ref{ryo-03}) of two-dimensional YO system.
Finally, due to the gauge freedom of $\tau$ function, we have the rational solutions to the multicomponent YO system (\ref{ryo-01}) in Theorem 1.


\end{document}